\documentclass[a4paper,superscriptaddress,twocolumn]{revtex4-1}
\usepackage{gensymb}
\usepackage{graphicx,color}
\usepackage{multirow}
\usepackage{times}
\usepackage{bm}
\usepackage{tabularx}
\usepackage{epstopdf}
\usepackage{enumerate}
\usepackage[normalem]{ulem}
\usepackage{verbatim}
\begin{document}

\title{An equation-of-state-meter for CBM using PointNet}
\author{Manjunath Omana Kuttan }
\email{manjunath@fias.uni-frankfurt.de}
\affiliation{Frankfurt Institute for Advanced Studies, 
D-60438 Frankfurt am Main, Germany}
\affiliation{Institut f\"ur Theoretische Physik,
 Johann Wolfgang Goethe Universit\"at, D-60438 Frankfurt am Main, Germany}
\affiliation{Xidian-FIAS international Joint Research Center, Giersch Science Center,
D-60438 Frankfurt am Main, Germany}

\author{Kai Zhou}
\email{zhou@fias.uni-frankfurt.de}
\affiliation{Frankfurt Institute for Advanced Studies, 
D-60438 Frankfurt am Main, Germany}

\author{Jan Steinheimer}
\email{steinheimer@fias.uni-frankfurt.de}
\affiliation{Frankfurt Institute for Advanced Studies, 
D-60438 Frankfurt am Main, Germany}

\author{Andreas Redelbach}
\email{redelbach@compeng.uni-frankfurt.de}
\affiliation{Frankfurt Institute for Advanced Studies, 
D-60438 Frankfurt am Main, Germany}
\affiliation{Institut f\"ur Informatik, Johann Wolfgang Goethe Universit\"at, D-60438 Frankfurt am Main, Germany}

\author{Horst Stoecker}
\email{stoecker@fias.uni-frankfurt.de}
\affiliation{Frankfurt Institute for Advanced Studies, 
D-60438 Frankfurt am Main, Germany}
\affiliation{Institut f\"ur Theoretische Physik,
 Johann Wolfgang Goethe Universit\"at, D-60438 Frankfurt am Main, Germany}
\affiliation{GSI Helmholtzzentrum f\"ur Schwerionenforschung GmbH, D-64291
Darmstadt, Germany}

\date{\today}

\begin{abstract}
A novel method for identifying the nature of QCD transitions in heavy-ion collision experiments is introduced. PointNet based Deep Learning (DL) models are developed to classify the equation of state (EoS) that drives the hydrodynamic evolution of the system created in Au-Au collisions at 10 AGeV. The DL models were trained and evaluated in different hypothetical experimental situations. A decreased performance is observed when more realistic experimental effects (acceptance cuts and decreased resolutions) are taken into account. It is shown that the performance can be improved by combining multiple events to make predictions. The PointNet based models trained on the reconstructed tracks of charged particles from the CBM detector simulation discriminate a crossover transition from a first order phase transition with an accuracy of up to 99.8\%. The models were subjected to several tests to evaluate the dependence of its performance on the centrality of the collisions and physical parameters of fluid dynamic simulations. The models are shown to work in a broad range of centralities (b=0-7 fm). However, the performance is found to improve for central collisions (b=0-3 fm). There is a drop in the performance when the model parameters lead to reduced duration of the fluid dynamic evolution or when less fraction of the medium undergoes the transition. These effects are due to the limitations of the underlying physics and the DL models are shown to be superior in its discrimination performance in comparison to conventional mean observables.  
\end{abstract}
\maketitle
\flushbottom
\section{Introduction}
Relativistic heavy-ion collisions produce small systems of strongly interacting matter of extremely high energy densities in which possibly a new state of deconfined matter, consisting of free quarks and gluons, called the Quark Gluon Plasma (QGP) is created \cite{Stoecker:1986ci}. The created hot and dense system expands rapidly under its own pressure and gradually cools down back to a dilute gas of hadrons which can be detected as final state particles in experiments. 
In QCD thermodynamics, the transition from a gas of hadrons to a QGP is likely a smooth crossover at high temperatures and very small baryon densities as established by lattice QCD \cite{Aoki:2006we,Borsanyi:2013bia,Bazavov:2014pvz}. A first order phase transition is conjectured at lower temperatures and moderate baryon densities \cite{Fukushima:2010bq}. Re-constructing the complete QCD phase diagram and identifying the regions of these transitions and thereby identifying the possible critical point, by means of experimental observations, is the major goal for the heavy-ion collision programs at Relativistic Heavy Ion Collider (RHIC), Large Hadron Collider (LHC) and the future Facility for Antiproton and Ion research (FAIR).
 
The Compressed Baryonic Matter (CBM) experiment at FAIR is a fixed target experiment that will study the phase structure of dense QCD matter with nucleus-nucleus collisions of energies up to $45 A$GeV in the lab frame \cite{Friese:2006dj, Senger:2006wd, Staszel:2010zza}. The physics program of the CBM experiment includes the exploration of high density equation of states such as in neutron star cores and the search for phase transitions at finite baryon densities \cite{Ablyazimov:2017guv,Senger:2020pzs}. The experiment will run at an unprecedented interaction rate of up to 10 MHz and the CBM detector will measure up to 1000 charged particles per collision. An online event selection algorithm \cite{deCuveland:2011zz} that performs ultra fast event reconstruction will be used to select interesting events for permanent storage from about 1 TBytes/s of collected data. Extracting the physics hidden in the vast amounts of data generated in this ambitious experiment requires the development of new techniques that can perform fast, accurate and real time physics analyses on raw experimental output. 

The incoming data stream from the detector is processed by different algorithms to  perform event reconstruction \cite{Kisel:2006yu}, particle identification and event selection \cite{deCuveland:2011zz} before different physics analyses can be performed. Events reconstructed and selected by these algorithms are used to calculate observables such as anisotropic flow and particle multiplicity fluctuations which are sensitive to a phase transition \cite{Rischke:1995pe}. Multi-parameter fits of the model simulations to the experimental data for these observables are currently used in experiments to search for phase transitions and to calculate the bulk properties of QCD medium. Bayesian analysis methods have been proposed as a method to fit the parameters to these observables\cite{Pratt:2015zsa, Bernhard:2016tnd,Bernhard:2019bmu}. An alternate approach to identify the appearance of a phase transition in QCD matter is based on Deep Learning \cite{lecun2015deep} techniques. Such DL techniques are considered so-called end-to-end approaches, where the DL model themselves determine the interesting features of the data and perform a classification task on these features. In \cite{Pang:2016vdc}, Convolutional Neural Networks \cite{gu2018recent} were trained on pion spectra (p$_t$, $\phi$) from hydrodynamic simulations to classify the EoS of a possible QCD transition. The study performed on the hydrodynamic output showed an average prediction accuracy greater than 95$\%$. A follow up study was presented in \cite{Du:2019civ} where a hadronic cascade model was employed after hydrodynamic evolution in the simulations to achieve a realistic freeze-out as well as including the effect of having a finite number of measurable particles in single events. The hadronic cascade "after-burner" introduces uncertainties in the final state spectra due to resonance decays and hadron rescatterings. This results in discrete particle spectra with predominant event-by-event fluctuations unlike the smooth spectra produced by pure hydrodynamic simulations. 
DL methods are reliable and accurate in identifying QCD transitions in heavy-ion collisions. However, as reported in \cite{Du:2019civ}, the performance depends largely on the fluctuations in the final state spectra. Therefore, if such a DL based EoS-meter is to be used on the direct output of a heavy ion experiment, an extensive analysis on the response of the DL model on additional uncertainties introduced by e.g. the detector resolution, acceptance region and efficiency of the reconstruction algorithms is necessary. The model should not only be robust against these constraints but also meet the performance in terms of accuracy and speed as demanded by the experiment.

In this study, the effects of experimental uncertainties and detector effects on the predictions of DL models for classifying QCD transitions at CBM experiment are investigated. The DL models were trained on a data similar to an experimental output by the use of a comprehensive data preparation pipeline that includes detector simulation and reconstruction algorithms. We demonstrate a novel DL model that can identify the EoS of QCD transition from raw experimental output and its performance on different situations of detector resolution and acceptance. We also studied its dependence on collision centrality and the model parameters for hydrodynamic evolution. This simulation study thereby shows for the first time how DL models can be employed in heavy-ion collision experiments to identify phase transitions directly from experimental output.

\section{The CBM detector}\label{cbmdet}
 The CBM detector is designed to make fast and precise measurements of the hadrons, muons and electrons produced in nucleus-nucleus collisions. The experiment will exploit modern radiation hard detectors with self triggered read out electronics to achieve the desired performance. Among the key components of the CBM experiment are the Silicon Tracking System (STS)\cite{heuser2013technical} and Micro Vertex Detector (MVD)\cite{Deveaux:2014cda} which are placed inside a superconducting dipole magnet with a magnetic field integral of 1 Tm. The MVD consists of 4 layers of Monolithic Active Pixel Sensors (MAPS)  placed 5-20 cm downstream the target. The main purpose of the MVD is to reconstruct open charm decay vertices and has an excellent position resolution of 3.5 - 6 $\mu m$ and secondary vertex resolution of about 50 $\mu m$. The STS comprises of 8 layers of silicon microstrip sensors placed 30 - 100 cm downstream the target. The task of STS is to reconstruct the tracks and momenta of charged particles. The STS has an excellent momentum resolution of about 1\%. Other sub detector systems of CBM include Ring Imaging Cherenkov Detector (RICH), a MUon CHamber system (MUCH), Transition Radiation Detector (TRD), Multi Gap Resistive Plate Chambers (MRPC) based Time of Flight (TOF) system, Electromagnetic CALorimeter (ECAL) and Projectile Spectator Detector (PSD). However, in this study we consider the data only from STS and MVD for the analyses.

\section{Microscopic and macroscopic dynamical models used to generate the data}\label{data prep}
To generate the training data for the DL analysis, this study uses the hybrid mode \cite{Petersen:2008dd} of the Ultra-relativistic Quantum Molecular Dynamics model (UrQMD 3.4) \cite{Bass:1998ca,Bleicher:1999xi} to simulate heavy-ion collision events with and without a phase transition. In this hybrid approach, a combination of microscopic and macroscopic description of collisions is used where the microscopic UrQMD model is used to generate realistic initial states of the collision at high baryon density. The consecutive hydrodynamic evolution models the intermediate hot and dense stage during which the system may undergo a phase transition \cite{Steinheimer:2007iy}. The hydrodynamic evolution starts once the Lorentz-contracted nuclei have passed through each other. This time ($t_{start}$) is given in natural units by
\begin{equation}
t_{start}=2R\sqrt{\frac{2m}{E_{lab}}}    
\end{equation}
where $R$ is the radius of the nuclei, $m$ is the mass of the nucleon and $E_{lab}$ is the kinetic energy of beam. At this time the particle list of UrQMD is transformed into an initial distribution of the energy-momentum and net baryon number density required for the subsequent hydrodynamic evolution. The required smoothing of the density is achieved by treating each hadron from UrQMD as a three dimensional Gaussian distribution of its energy-momentum as well as baryon number. One should note that this initial state will give reasonable event-by-event fluctuations for the initial eccentricities and is also independent of the equation of state that is employed for the fluid dynamical evolution. Any effect of the EoS will therefore be confined only to the expansion phase. The SHASTA \cite{Rischke:1995ir,Rischke:1995mt} algorithm is then used for the 3+1D ideal fluid dynamic evolution on a Cartesian grid with a spacing of $\Delta x= 0.2$ fm and a grid size of $200^3$ cells.

\begin{figure}
    \centering
    \includegraphics[width=0.5\textwidth]{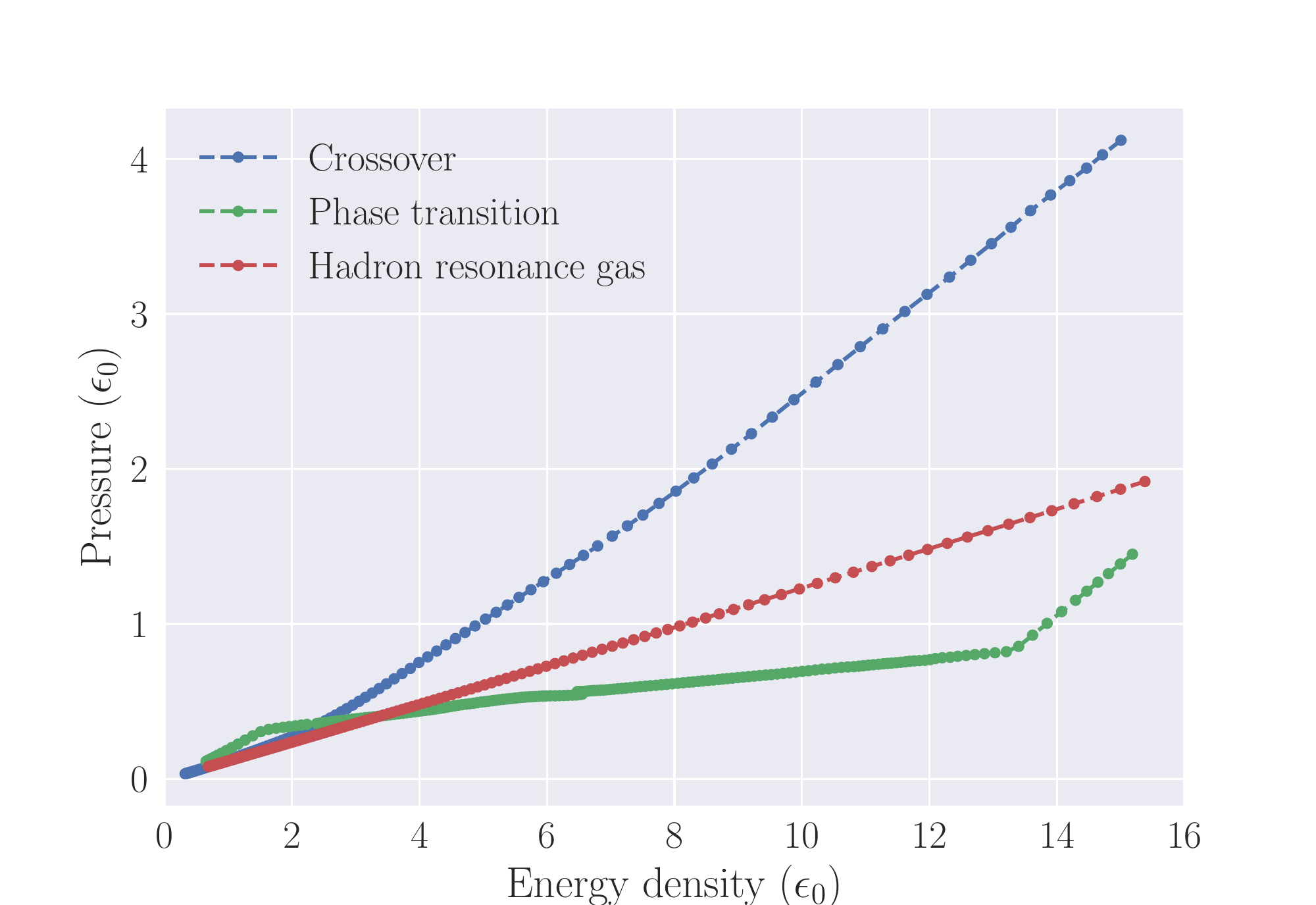}
    \caption{(Color online) The Equations of State, along an isentropic trajectory, for first order phase transition, crossover transition and  hadron resonance gas as incorporated in the simulation. The pressure of the medium for a central cell is plotted as a function of its energy density in central Au-Au collision at $10 \ A$GeV in lab frame. The phase transition  is associated with a plateau region where the pressure remains almost constant as the system evolves while the pressure is a continuous function for a crossover transition. The phase transition and crossover EoS are used to train the models while the hadron gas EoS is used only to test the performance of the models on an unseen EoS. }   
    \label{0}
\end{figure}

The equation of state of the medium is an essential input that is required to solve the fluid dynamic equations. The EoS combines the microscopic and macroscopic properties of the system created and provides the pressure of the medium for any given energy and net baryon number densities. The EoS incorporates the QCD transition, as the evolution of the medium is driven by pressure gradients. In this study, we use two distinctly different equations of state for training and validation. One based on a Maxwell construction between a bag model quark gluon EoS and a gas of pions and nucleons \cite{Rischke:1995ir,Rischke:1995mt} to simulate the first order phase transitions scenario. The second EoS is dubbed the Chiral Mean Field hadron-quark EoS \cite{Steinheimer:2010ib} which describes a smooth crossover transitions as predicted by lattice QCD. To investigate the models output for an unknown EoS we also employ a hadron resonance gas  equation of state which is based on a free gas of hadrons according to the known hadronic resonances from the particle data group \cite{Zyla:2020zbs}. The three equations of state, along trajectories of constant entropy per baryon, as expected for heavy ion collisions at $\mathrm{E_{lab}}=10 \ A$GeV, are visualised in figure \ref{0}. While the crossover EoS is the stiffest and the phase transition the softest equation of state, the HRG lies in between these two extreme cases.

 The fluid dynamical evolution proceeds until the energy density in all cells falls below a freezout energy density ($\epsilon$) after which the evolution is stopped. The default value for $\epsilon$ is five times nuclear ground state energy density $(\epsilon_0)$ but it can be adjusted freely. More details on the motivations behind the chosen values for  $t_{start}$ and $\epsilon$ are discussed in \cite{Petersen:2008dd,Steinheimer:2007iy}. Particles are then generated from an iso-energy density hypersurface which has been created throughout the whole time evolution. The density that defines this particlization hypersurface is the above defined value of $n \epsilon_0$. The sampling of particles is done using the well known Cooper-Frye formula
\begin{equation}
E\frac{dN}{d^{3}p}=\int_{\sigma}f(x,p)p^{\mu}d\sigma_{\mu }
\end{equation}
where $f(x,p)$ is the boosted Fermi or Bose distribution and $d\sigma_{\mu }$is the freeze out hypersurface element. Here, global conservation of baryon number, charge, strangeness is exactly observed.
The particles are then transferred to UrQMD where hadronic cascade calculations happen. Important final state effects such as hadronic rescattering and resonance decays are performed at this stage.
The output of the UrQMD-hybrid model is then an event-wise list of particles with their four momenta and positions.

The main objective of the study is to develop a DL model that uses information similar to the experimental output of the CBM experiment, without any significant analysis chain. Furthermore our study will analyse the effects of experimental constraints on the performance of this model. Therefore, an accurate modelling of the experimental condition is necessary. The CbmRoot \cite{root_url} package is used to transport the final state particles from UrQMD through the CBM detector simulation. CbmRoot uses GEANT3 \cite{Brun:1987ma}  to simulate the electromagnetic and weak interactions as well as decays of particles traversing the detector. The hits in the detector are then digitised to mimic the detector resolution and finally these digitised hit positions are used to reconstruct the tracks using a Kalman filter based algorithm \cite{Kisel:2006yu}. The standard CbmRoot macros are used for the transport simulation, digitisation and track reconstruction. As a result we obtain realistic event-wise output from the detector simulation which now can be used as input for the DL analysis.

It is also important to note that CbmRoot can perform the full detector simulation according to the experimental specifications. However the default setup does not include a realistic simulation of different backgrounds which may lead to additional noise and could potentially weaken the discrimination performance. 
In the actual experimental data taking, quasi real-time processing of free-streaming detector data requires an extra stage of event building, i. e. the identification of clusters of detector hits sufficiently close in time and space. After the step of event building separate events are technically defined and can be processed, also in the approach of this analysis. 
It is interesting to note that the process of event building might also be improved by DL-based methods, similar to the PointNet recently developed in \cite{Kuttan:2020kha}.

\section{PointNet for classifying the EoS}
Deep Learning is a well established Machine Learning method inspired by the way information is processed in biological systems. It employs multiple layered Artificial Neural Networks to learn higher dimensional correlations in the data. Machine learning and Deep Learning methods have been widely used both in theory \cite{Zhou:2018ill,Fujimoto:2019hxv, Steinheimer:2019iso, Thaprasop:2020mzp,Pang:2019aqb,Wang:2020hji,Jiang:2021gsw,Shi:2021qri,Song:2021rmm, Li:2020qqn, Wang:2020tgb, Kvasiuk:2020izb, Boyda:2020hsi, Liu:2020omw,Pang:2021vwl} and in experimental high energy physics \cite{Bourilkov:2019yoi,Radovic:2018dip,Guest:2018yhq,Larkoski:2017jix,deOliveira:2015xxd,Baldi:2016fql,Komiske:2016rsd,Almeida:2015jua,Kasieczka:2017nvn,Kasieczka:2019dbj,Qu:2019gqs,Moreno:2019bmu,Samuel:2018xci,Samuel:2019crc,Kasieczka:2020nyd,Sirunyan:2020lcu,Esmail:2019ypk,Haake:2017dpr,Samuel:2021jho,Banerjee:2020iab}. Previous studies \cite{Pang:2016vdc,Du:2019civ} on identifying the QCD phase transitions have shown that Convolutional Neural Network (CNN) based models can accurately classify the underlying equation of state from a hydrodynamic evolution using the p$_t$- $\phi$ spectra of pions (differential transverse and angular distributions in the transverse plane). In \cite{Sergeev:2020fir}, CNN was used to detect the formation of QGP in CBM experiment. CNNs are a good choice of algorithm for extracting correlations from image like data, i.e. data which is provided in the form of equally spaced multi dimensional histograms. However, the purpose of this study was to train DL models directly on experimental outputs such as the information of discrete reconstructed tracks of particles in a collision event. The state vector which represents a reconstructed track in a CBM detector plane comprises of transverse x, y coordinates, tangential directions to the track and the charge to momentum ratio (q/P) of the particle. This data can be fed to a neural network as a 3D voxel array (trajectories in 3D) or as two separate 2D pixel arrays (trajectories in x-z and y-z planes). However, this would render the data to be highly voluminous causing large memory requirements. Moreover, processing the data into these images and combing through it with CNNs would be computationally inefficient and slow. Considering the potential use of a DL based EoS meter for fast online data analysis at CBM, this conversion of the data to images could slow down the whole analysis chain. A solution to these issues is to use a point cloud representation of the data. Point clouds are collections of disordered points in space. A track can be considered as a point in the point cloud in a N-dimensional space where "N" is the number of attributes describing the track. The data therefore becomes an order invariant list of tracks where each entry of the list is the state vector of the track.

DL models can be trained on point cloud data using the PointNet \cite{qi2017pointnet} architecture. PointNet based DL models have been shown to learn from heavy ion collision data to reconstruct the impact parameter of collisions in \cite{Kuttan:2020kha, Kuttan:2021zcu}. In this study, we used a similar network architecture but less complex (i.e.; lesser number of trainable parameters) than the one described in the above paper. The PointNet based models accept the point cloud in the form of a 2D array where each row is a point (i.e. a track information in the event) in the point cloud and each column is an attribute of the point/track. This array is then processed with symmetric, order invariant operations to extract global features which finally pass through a fully connected deep neural network to identify the EoS that created the given point cloud.

\section{Training and testing PointNet models}
The present study was conducted on a set of Au+Au collisions at a beam energy of $10$ $A$GeV in the lab frame. CBM will also study other heavy ions at similar energies. However, as the underlying physics of the collisions remains the same, the models developed in this study can be easily extended for application to other nuclei. The dataset for this study was generated using the UrQMD-hybrid model and CbmRoot package as described in section \ref{data prep}. It consists of 30000 training events and 10000 validation events each for the crossover and first order phase transition equation of states with an uniform impact parameter (b) distribution from 0 to 7 fm.  To study the effects of experimental uncertainties and constraints on the performance of the DL models, the PointNet model was trained on different outputs: 
\begin{enumerate}
\item Firstly, the final state output (\emph{Dataset 1}), i.e. the particle information directly from the UrQMD model without any acceptance cuts. This dataset contains essentially the full event information and has not been transported through the detector simulation. 

\item Secondly, the final state output within CBM detector acceptance (\emph{Dataset 2}). The dataset contains final state particles from UrQMD model within the CBM acceptance region of 2-25$\degree$ polar angles. This corresponds to a hypothetical, ideal detector output which detects all particles within its acceptance with infinite resolution.

\item Lastly, the CbmRoot simulated data (\emph{Dataset 3}), i.e. the final state output from UrQMD is passed through CbmRoot. This dataset comprises of the reconstructed tracks from the digitised hits of particles in the simulated CBM detector.
\end{enumerate}

The network structure and other training parameters were fine tuned through trial and error to achieve the best performance on the final state output (\emph{Dataset 1}). The same network architecture and hyperparameters (however, with different input dimensions depending on the dataset) were then used for training the model with experimental effects (\emph{Dataset 2,3}). In this way, it was possible to study the response of the same DL network to different experimental constraints. 

\subsection{Network architecture}
The input point cloud passes through 3 1D-convolution layers to extract 128, 256 and 512 feature maps respectively. Batch normalisation layers are present between every convolution layer. An average pooling layer then extracts one global feature of the point cloud from each of the 512 feature map generated by the final convolution layer. The 512 global features are the input to a 3 layer fully connected Deep Neural Network (DNN) with 256, 128 and 2 neurons respectively. Batch normalisation and dropout layers (with drop out probability 0.5) are present between every DNN layer. All layers except the final layer use the ReLU activation function. A softmax activation is used on the final layer to classify the EoS. The models use the Adam optimiser with a learning rate of $10 ^{-5}$ and categorical cross entropy as the loss function. The models were trained until the network started overfitting the data and the best model in terms of validation accuracy and loss was  chosen for further analyses.

\subsection{Training results}
\begin{figure}
    \centering
    \includegraphics[width=0.5\textwidth]{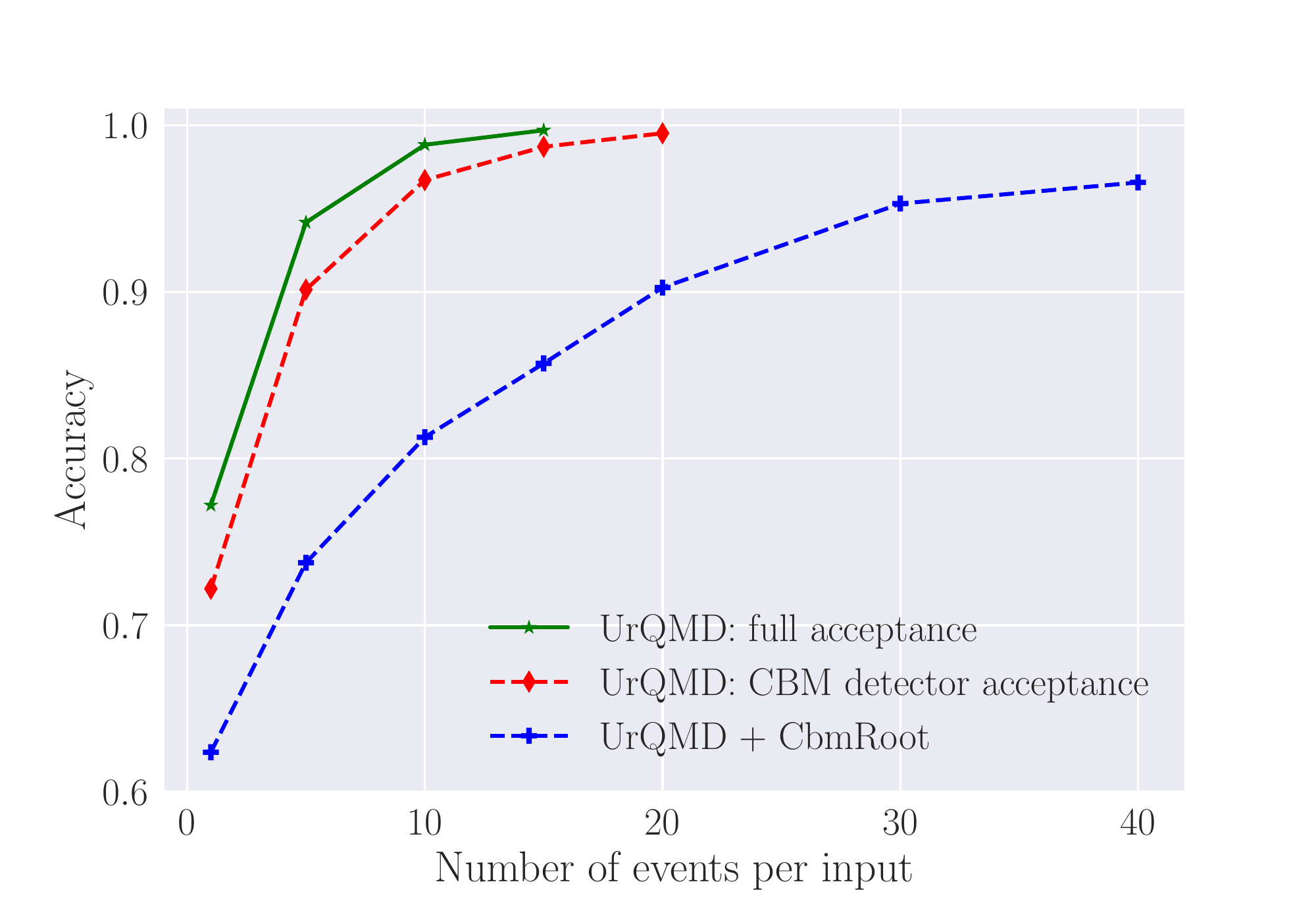}
    \caption{(Color online) Validation accuracy of the PointNet models as a function of number of  events combined to create the input. The number of training and validation samples were fixed to be 60000 and 20000 respectively for all the models although multiple events were combined to create a sample. While randomly combining events, it was ensured that an event in training sample was never present in a validation sample. The DL models achieved greater than 99\% accuracy with the combination of 15 events in an ideal case while 20 events had to be combined in the presence of an acceptance cut. However, the DL model required a combination of 40 events to achieve about 96\% accuracy in a more realistic experimental condition. }
    \label{1}
\end{figure}

As discussed above, three different scenarios for the input data were investigated in this study. In the first case (\emph{Dataset 1}), the input for training was the event-by-event list of four-momenta of all particles from UrQMD. The input data has dimensions N$\times 4$ where N is the maximum number of particles present in an event. Events with less number of particles are filled with zeros to maintain the same input dimensions. In this scenario, the trained PointNet model achieved a validation accuracy of 77.2\% for the correct event-wise classification between crossover and phase transition EoS. This accuracy can be improved if multiple events are combined to create the input. This was done by randomly selecting K events, i.e. all rows (without replacement) in that event, from the event-by-event lists (along with rows filled with zeros) and concatenating them to create a longer list with dimensions (K*N)$\times 4$. It must be noted that the combined events are randomly chosen from b=0-7 fm. A validation accuracy of 99.7\% was achieved by the model when the input was the combined data from 15 events as can be seen in figure \ref{1}. The model learns a set of unique observables for classifying the underlying EoS and the boundaries of these observables for either classes are accurately learned with a combined dataset.

In the second case, the input for training was the four momentum of particles from UrQMD which were within the CBM detector acceptance. Particles beyond the CBM acceptance range of a 2-25$\degree$ polar angles
were removed from the events. The validation accuracy in this case was decreased to about 72.2\% for the event by event input and the model was able to achieve an accuracy of 99.5\% by combining 20 events for the input.

The decrease in accuracy can be understood. Supplying the PointNet with only a shortened or partial list of particles increases the difficulty of learning the observables capable of classifying the EoS. The DL model therefore requires a few more events to achieve a classificaton accuracy similar to the first case. The models cannot distinguish particles belonging to one event from another. Therefore, it is likely that the unique DL constructed observables are some aggregate quantities, probably within certain region of the phase space. An acceptance cut could remove part of the information which was otherwise available (in first case) and calculating these observables accurately would naturally require more statistics.  

In the third dataset, more realistic experimental constraints of acceptance and resolution were introduced. The UrQMD output was passed through the CBM detector simulation and the model was trained on the tracks reconstructed from the hits of the particles in MVD and STS detectors of the detector simulation. In this case, the average classification accuracy for single event inputs was only 62.4\%. However, after combining 40 events for an input, the accuracy increased again to 96.6\%. For this model to achieve a performance similar to the second dataset, the number of events that were combined to create the input had to be doubled. This model, based on dataset 3, that uses 40 events of reconstructed tracks as input is henceforth referred to as \emph{Model-1}. 

The accuracy of PointNet models in the three cases as a function of number of events combined is plotted in figure \ref{1}. It is evident from the plot that the performance of DL model is only marginally decreased in the presence of a simple acceptance cut but there is a large drop in the accuracy when a more realistic experimental scenario is considered. This shows that the final state particles have strong features that are characteristic of the macroscopic properties that governed the evolution of QCD medium. However, in an experiment these distinct features become weaker and difficult to identify. Uncertainties in measurements due to the detector resolution and randomness in the detected particle spectra arising from interactions of particles in the detector diminish the relevant signals in the data. Inefficiencies of reconstruction algorithms and selection cuts also introduce errors in the final data. Nevertheless, the DL model is able circumvent these issues by combining more events for decision making. A similar behaviour was also reported in \cite{Du:2019civ}. Increasing the statistics  reduces the stochasticity in the data thereby improving the predictive power of DL. For instance, the global feature used by the PointNet models for classifying EoS are the average (given by average pooling layer) values of each feature extracted by the convolution kernels. These averages could be more accurate determined when more sample points are used. In this way, the PointNet models could improve in performance when more events are used.

\begin{figure}
    \includegraphics[width=0.5\textwidth]{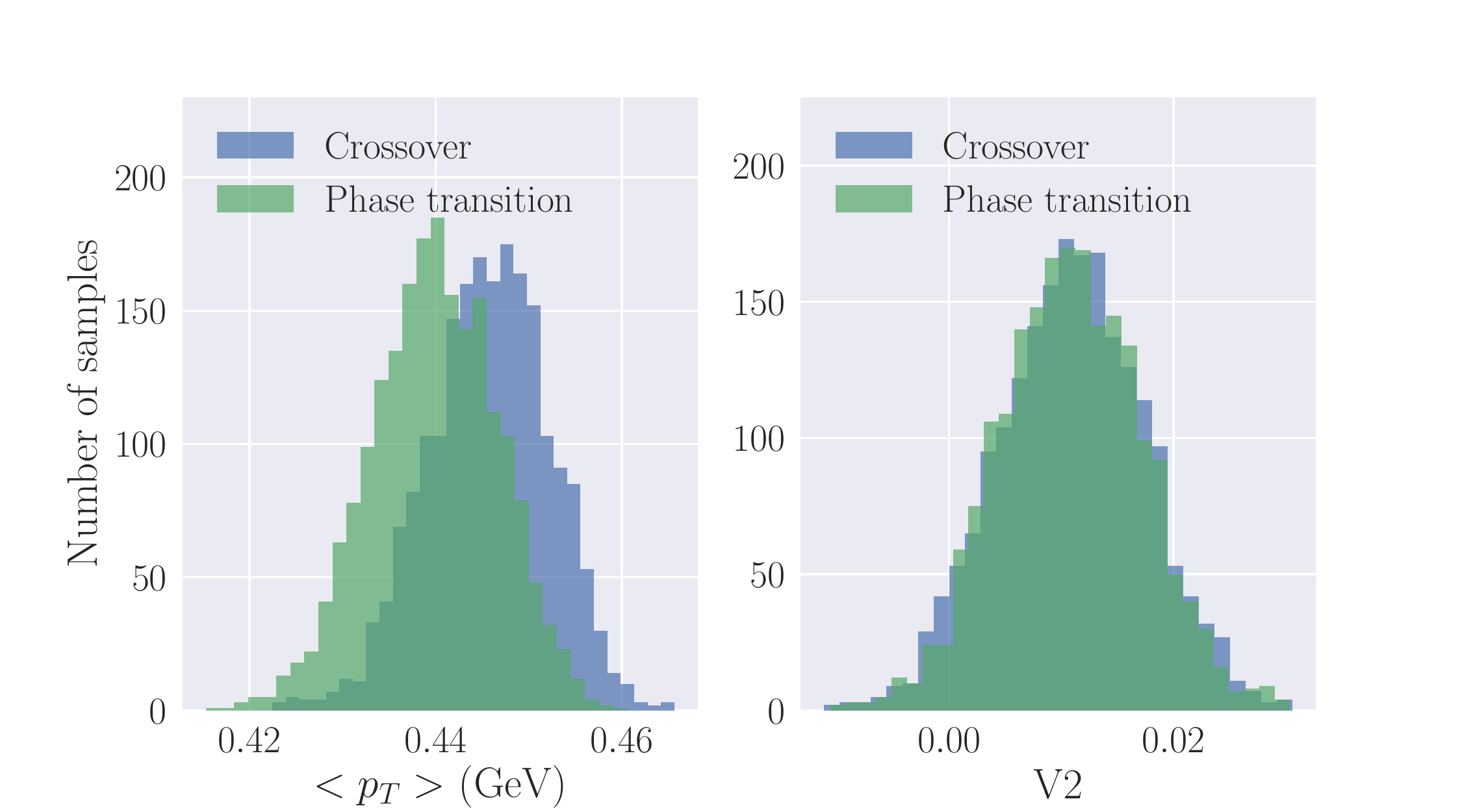}
    
    \caption{(Color online) Distributions of Mean transverse momentum  (left) and elliptical flow (right) for crossover and first order phase transitions. The values are averaged over all particles from 15 UrQMD events with b=0-7 fm (\emph{Dataset-1}). The distributions have significant overlap that it is not possible to classify the EoS using these observables while the DL model with 15 event combined input achieved an accuracy of 99.7\%.}
    \label{ptv}
\end{figure}

However, this does not mean that conventional mean observables such as mean transverse momentum ($<p_{T}>$), collective flow ($v2$) etc. can be used for classifying the EoS as accurately as PointNet models. The above mentioned DL models do not require any event selection based on centrality while the traditional observables have strong centrality dependence. Without a centrality selection and high statistics, the traditional observables will not have well separated boundaries that can aid an accurate classification. The $<p_{T}>$ and $v2$ distributions for 15 events averaged data from UrQMD are plotted in figure \ref{ptv}. It is evident from the plots that the distributions of these observables, after averaging over only 15 events, overlap significantly and cannot be used to classify the two classes of data as accurately as the PointNet model does. We have also checked that simply calculating averages of the different components of the input features in the PointNet will also not lead to easily distinguishable distributions. A more in depth discussion on the interpretability of the network is given in appendix \ref{appendix}. There, we describe a method to interpret the decision making process of our PointNet model which helps to understand why the model outperforms conventional observables.

In other words, the PointNet model is able to learn unique  observables that produce a close to perfect classification accuracy from only combining the input of 15 random events. The PointNet model is able to learn such observables even from an "experiment like data" in which the reconstructed tracks are the input (\emph{Model-1}).

In order for the CBM experiment to make complete utilisation of the high event rates, accurate online event selection and analysis techniques are necessary. The DL models require a maximum of just 40 events to achieve a classification accuracy greater than 96\%. The PointNet based EoS meter can serve  this purpose and can be coupled with other DL based algorithms (eg. centrality meter \cite{Kuttan:2020kha}) for a comprehensive online event analysis. 

It is well known that conventional observables are very sensitive on model parameters such as the centrality selection, initial state, freeze-out condition etc. Therefore, we investigate in detail the generalisation ability of the PointNet models on these parameters in the following sections.

\subsection{Centrality dependence}
\emph{Model-1} which had an accuracy of 96.6\% was trained on events with impact parameters 0-7 fm. Although the accuracy is already good enough, the model showed slightly better performance on central events which hints to a centrality dependence. To examine if the accuracy of the model can be increased with a different centrality selection, a model (\emph{Model-2}) was trained explicitly on events with an impact parameter of 0-3 fm. This model also used the tracks reconstructed from the detector and combined the data from 40 events to form an input. The \emph{Model-2} achieved a prediction accuracy of 99.8\% on events with impact parameters 0-3 fm: Choosing a smaller centrality bin therefore improved the performance of the model. However, most of the events collected in the experiment will be unusable if we choose only central collisions. To tackle this issue, a model (\emph{Model-3}) was trained which combined only events with impact parameters 0-3 fm and 3-7 fm separately. The input for this model was a combinations of 40 events (reconstructed tracks) either from the impact parameter bin of 0-3 fm or from 3-7 fm. In addition to this selection of events, the network had 1 extra input to feed in the impact parameter bin of the given sample (i.e; 0 if b=0-3 fm and 1 if b=3-7 fm). This input is concatenated with other extracted global features and is fed into the DNN. The \emph{Model-3} achieved a validation accuracy of about 99.65\% for events with impact parameter 0-3 fm and 81.27\% for impact parameter 3-7 fm. The PointNet models can achieve the best performance for central events, assuming they can be accurately identified \cite{Kuttan:2020kha}. However, significant accuracies can also be achieved on peripheral events if they are separated from central events for training.

\subsection{Dependence on model parameters}
\begin{figure}[t]
    \centering
    \includegraphics[width=0.5\textwidth]{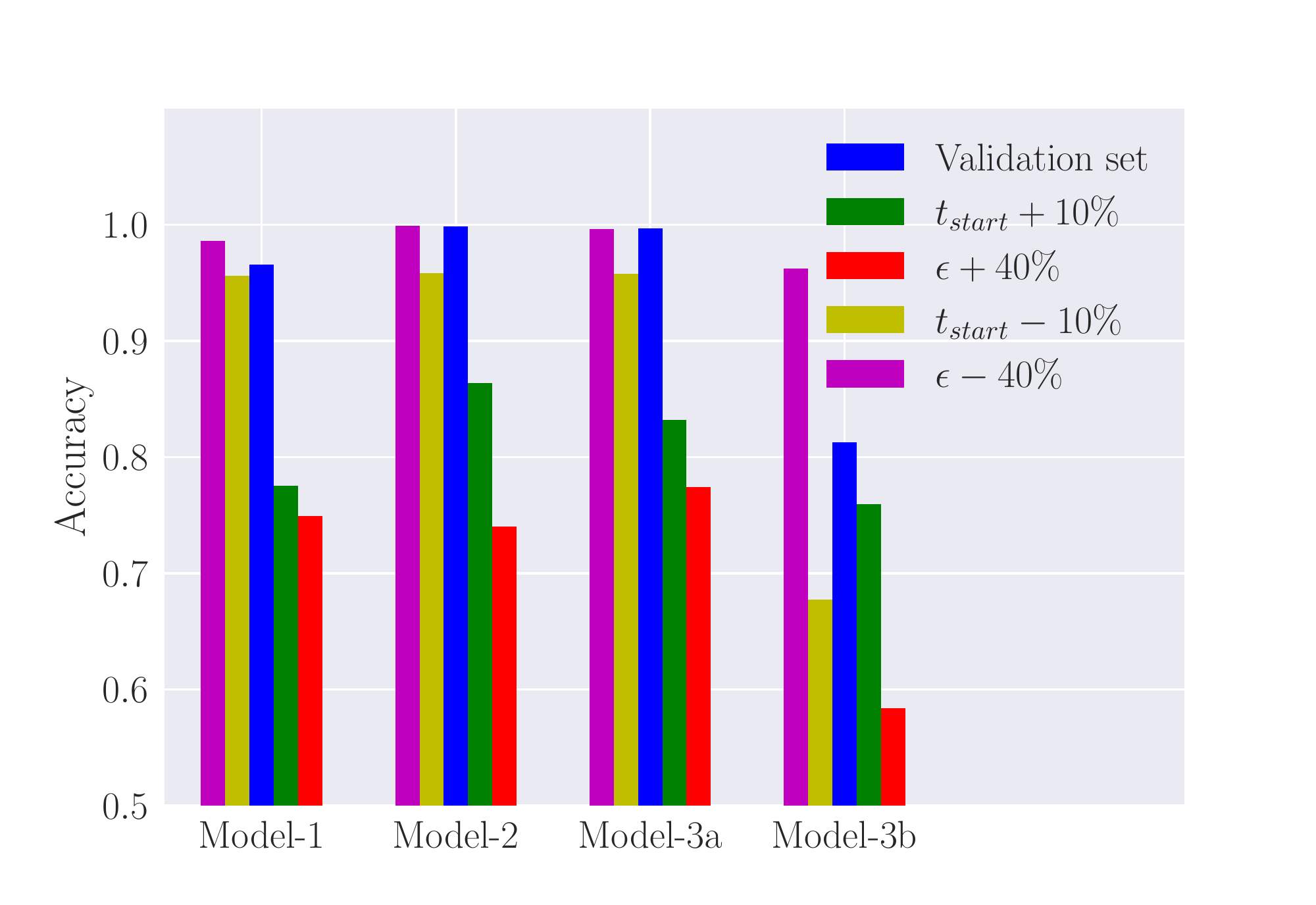}
    \caption{(Color online) Variation in the testing accuracy of the PointNet models with change in $t_{start}$ and $\epsilon$. The blue bars show the validation accuracies of the models while the other colours represent the testing accuracy on datasets different from the training data. Each testing dataset comprised of 2000 events for each EoS. The \emph{Model-3a} and \emph{Model-3b} are the testing results of the \emph{Model-3} on impact parameters 0-3 fm and 3-7 fm respectively.}
    \label{2}
\end{figure}

\begin{figure}[t]
    \centering
    \includegraphics[width=0.5\textwidth]{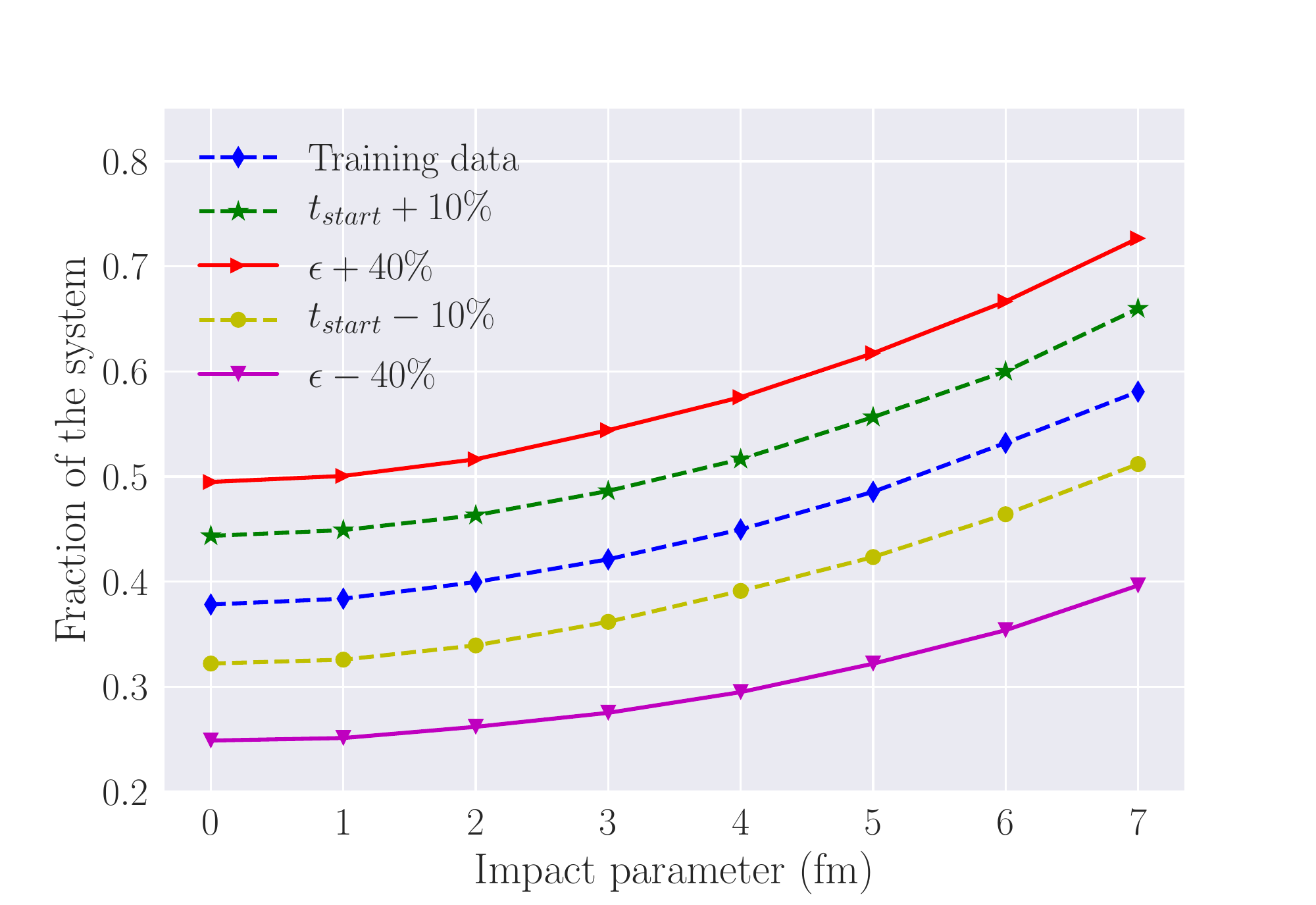}
    \caption{(Color online) Fraction of the medium which is below the freezout energy density at the beginning of hydrodynamics as a function of impact parameter. This is simply the fraction of the medium which does not undergo the hydrodynamic evolution. The blue curve represents the initial conditions used while training the model. Curves above the blue curve corresponds to the initial conditions which reduces the duration of hydrodynamic evolution and vice versa.}
    \label{fr}
\end{figure}

In the previous section, it was shown how PointNet models can be employed to correctly classify the nature of the QCD transition with large accuracy in a wide range of centralities or in a small centrality range depending on the experimental requirement. However, the physical and model parameters have been kept constant, i.e. they where assumed to be known exactly. In reality this is not the case and thus, to ensure the reliability of DL model in an experiment, the models must be robust against reasonable changes of the physical parameters of the hydrodynamic event generator. Two such parameters are the starting time for hydrodynamic evolution ($t_{start}$), which essentially determines the time at which one can assume local equilibration to be reached and the particlization energy density ($\epsilon$), which determines at which point the system starts to fall out of local equilibrium. Since at that energy density particles are emitted from the hydro to the non-equilibrium hadronic rescattering phase, matter below this criterion will effectively not be influenced by the EoS. To evaluate the dependency of the DL models on these parameters, the trained PointNet models were tested on events where $t_{start}$ is varied by $\pm$ 10\% and $\epsilon$ by $\pm$ 40\% from the training value. The performance of the DL models are illustrated and compared in figure \ref{2}. The models in general seem to achieve an accuracy similar to the validation accuracy when $t_{start}$ or $\epsilon$ is decreased. However, the accuracies decrease considerably when the $t_{start}$ or $\epsilon$ is increased. This effect can be understood if one studies the fraction of the matter which is below the particlization criterion, and therefore does not carry any information on the EoS, for the different initial and freeze out conditions . This fraction varies also as a function of the impact parameter as shown in figure \ref{fr}. 
The decrease in performance with decrease in the duration of hydrodynamic evolution and centrality is therefore nicely illustrated with figure \ref{fr}. It can be seen that a smaller portion of the emitted hadrons experiences the dynamics of the phase transition as the impact parameter is increased. This explains the higher validation accuracy for \emph{Model-2} compared to \emph{Model-1} and the decreased validation accuracy for \emph{Model-3} on events with impact parameter 3-7 fm.

A delayed starting time of the hydrodynamic evolution or an increased freeze out energy density reduces the contribution of the hydrodynamic evolution of the system to the emitted particles and therefore the EoS will have less influence on the final particle spectra. While an increase of the duration of the hydrodynamic evolution leads to a prolonged influence of the EoS on the evolution of the medium and thus a higher accuracy, the performance drop can be related to a limitation imposed by physics which may not be avoidable.

 Similarly, an increase in the freeze out energy density by 40\%, for b=0 fm, causes about 50 \% of the final particles being already emitted before the hydrodynamic evolution even begins. The DL-models have to rely on the artefacts left by the EoS in the remaining 50\% of the emitted particles to make a decision. This is why the accuracy decreases considerably with an increase in freeze out energy density. However, the decrease of the portion of the emitted particles that undergo the hydrodynamic evolution from an increase in the freeze out energy density by 40\% is larger than when $t_{start}$ is increased by 10\%. This is why the drop in the accuracy is comparatively lower when the $t_{start}$ is increased by 10\%. In short, hadrons from central events with early starting of the hydrodynamic evolution or a decreased freeze out energy density carry more information on the EoS as they are, on average, emitted after a longer hydrodynamic evolution.

\subsection{Testing on an unseen EoS}
We have shown that the PointNet models can accurately classify the data into one of the two training EoS. However, the actual EoS of the fluid dynamic evolution can be different from the ones used during the training. To understand how the DL model would perform in such a scenario, we tested \emph{Model-1} on an EoS which it was not trained on. On the hadron resonance gas EoS, the \emph{Model-1}  classified 68\% of the samples as crossover and the remaining as a first order phase transition. As evident from figure \ref{0}, the hadron gas EoS is similar to a crossover EoS. At low energy densities, the hadron gas EoS traces the crossover equation of state and at high densities, the pressure is in between the phase transition and crossover equation of states. The hadron gas EoS also doesn't have a plateau like region of constant pressure which is characteristic of the phase transition EoS. This explains why the model prefers to predict the hadron gas EoS as a crossover equation of state. The PointNet based binary classifier of EoS can therefore provide reliable predictions not just on the trained EoS  but also on other similar equation of states for crossover and phase transition.

\subsection{Comparison to a single event classifier}
We have shown that the performance of the model can be improved by combining multiple events to train the PointNet models. A recent study \cite{Nachman:2021yvi} pointed out, that a single event classifier, when applied on N events could outperform a classifier trained on combinations of N events if these events are statistically independent. This raises the question if an event-by-event EoS classifier, combined over N events, would outperform the combined events models developed in this study. To check this, 20000 validation events of \emph{Dataset-3} were tested using a model trained to classify the EoS of individual events. The final prediction is then defined as the predicted EoS of the majority of the events for groups of 40 random events. This procedure achieved an accuracy of 92.01\%. At the same time, the \emph{Model-1} which was trained on combinations of 40 events to make predictions had an accuracy of 96.6\%. The single event classifier therefore doesn't achieve the accuracies of the combined events classifier. An accuracy of about 92\% can be achieved by training the model on combinations of about 25 events while the single event classifier required 40 events to achieve the same accuracy. The superior performance of the PointNet models trained on combinations of multiple events is due to the centrality dependent influence of the EoS on the system. As shown in figure \ref{fr}, a significantly larger fraction of the system is influenced by the supplied EoS for a central event while most part of the system is not influenced by the EoS for peripheral events. Therefore central events contain more information on the EoS which governed its evolution than a peripheral event. When the PointNet is trained on combinations of random events with all centralities, the model can learn to make decisions using the signals from the central events present in the data. A single event classifier on the other hand would struggle to correctly classify the peripheral events which would often contain only very weak signatures of the EoS. This centrality dependent performance bias would further worsen the performance of single event classifiers when a realistic impact parameter distribution $(P(b) \propto b)$ is considered where the central events are rare compared to peripheral events. Another practical advantage of using combinations of events is that such models can potentially work with a continuous datastream without event building or event separation. This can be extremely useful to the CBM experiment which will require extremely fast analysis methods for the data collected at rates upto 10 MHz.   

\section{Conclusion and Discussion}\label{conc}
In this study, we have developed PointNet based DL models that can extract very complex universal event features from basic event information of heavy ion collisions at the CBM experiment. This model is even able to classify events by very abstract event features like the EoS present during the hot and dense stage of the collision, i.e. whether a phase transition was present or not. The prediction accuracy was found to be improving when more events were combined to make the predictions. This shows that with increased statistics, PointNet models learn the global features that can classify the EoS despite the uncertainties in the data arising from a discrete particle spectra with final state effects, detector effects and inefficiencies of reconstruction algorithms. It is noteworthy that the PointNet models can achieve a classification accuracy of up to 96.6\% from the reconstructed tracks of  particles from just 40 collision events. The PointNet models can work on a wide range of impact parameters but they achieve the best performance by choosing only central collisions for analysis. However it is also possible to include non central collisions for analysis if central collision events are not mixed with non central collisions. The predictions of the DL models were also robust to some changes in the physical parameters like the initial condition. The performance of the models was consistent when $t_{start}$ or $\epsilon$ was decreased from the training value while a decrease in the performance is observed when these parameters are increased. This is interpreted as a physical consequence of a decreased influence of the hydrodynamic evolution, and the EoS, on emitted particles. Nevertheless, the DL models show good performance in comparison to conventional averaged event features like  $<p_{T}>$ or $v_2$ which have similar values for both the classes. The values of these features also differ widely for different model parameters. 

The use of experimental output such as the tracks of particles can eliminate any possible biases in the data that might appear in later stages of data processing. The point cloud representation of data requires minimal pre-processing before being fed to the DL model. This enables the model to be deployed in the experiment for fast, online analysis of experimental data. Moreover, these models can be easily translated to any other heavy-ion collision experiment for similar tasks. The capability of these models to work on large range of impact parameters make it an ideal tool to search for phase transitions in  heavy-ion programmes. Due to their ability to find out global features in the input, the PointNet based models can also be easily adapted for analysing any other global event feature of heavy-ion collisions. Future studies in this direction can be focused on incorporating more equation of states making it a multi class classification problem and testing the performance of the models for other FAIR energies. It would also be interesting to study the performance of DL models in a continuous datastream and in the presence of detector noise, event pileup etc. Studies on training the DL models on low level detector data such as the signals from readout channels and  deploying them directly on the detectors using FPGAs is another interesting direction which could be extremely beneficial to the CBM experiment. Such methods can be exploited for ultra fast event selection and analysis based on yet unachievable, complex event features.

\begin{acknowledgments}
The authors thank Benjamin Nachman and Jesse Thaler for their insightful comments and discussions. M.O.K. thanks the GSI and HFHF as well as the SAMSON AG for their support.
 K.Z. and J.S. thank the Samson AG and the BMBF through the ErUM Data project for funding. H.S. acknowledges the Walter Greiner Gesellschaft zur F\"orderung der physikalischen Grundlagenforschung e.V. through the Judah M. Eisenberg Laureatus Chair at Goethe Universit\"at Frankfurt am Main.

\end{acknowledgments}

\bibliographystyle{apsrev}
\bibliography{eos.bib}
\clearpage
\appendix
\section{Interpreting the PointNet model}
\label{appendix}
It is generally interesting to reveal how the PointNet model is able to accurately discriminate the QCD transitions even under conditions where the conventional observables failed. However, interpreting the inner workings of a neural network with conventional concepts is not straight forward especially when the inputs are order invariant, as in our case. Nevertheless, in \cite{qi2017pointnet}, a method to visualise the critical points of a point cloud is discussed. The PointNet architecture comprises of several 1-D convolution layers followed by a symmetric function which converts each feature map produced by the last convolution layer into a single number. These numbers which are considered global features of the pointcloud form the input to a fully connected neural network (DNN) which classifies the input point cloud. In \cite{qi2017pointnet}, a maxpooling layer is used as the symmetric function to extract the global features. In other words, the feature which has the largest numerical value in each feature map given by the last convolution layer becomes the input to the fully connected neural network. Each of these features can be traced back to the original point in the point cloud. Such points are then defined as the critical points of the point cloud as they directly induce the input to the DNN which classifies the data.

We extended this method to analyse the decision process of our EoS classifier. Our model produces 512 global features for each point cloud. These global features are then used by a fully connected network to make the classification. Different to the above described method, the symmetric function used in our study to generate these features, is average pooling. This has severe consequences on the interpretability since the average of the feature map, given by the average pooling layer, cannot be uniquely traced back to a single point in the point cloud. However, we can still attempt to identify those features which seem most important for the classification task and then analyse which properties of the input point cloud affect these features.

\begin{figure}
    \centering
    \includegraphics[width=0.5\textwidth]{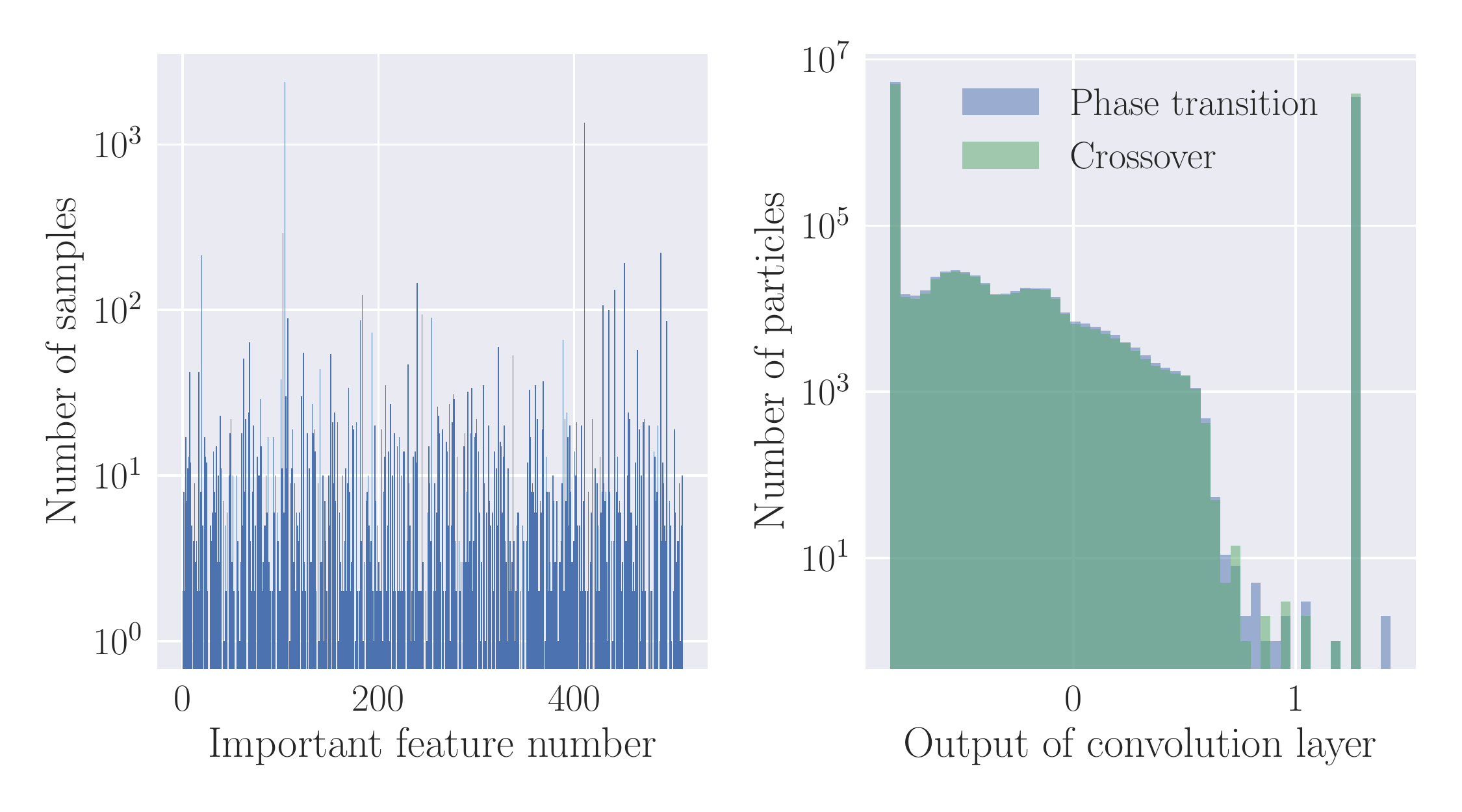}
    \caption{Left: Distribution of the feature number of the most important feature for 10000 pairs of events. An important feature is the global feature which has the largest difference in their values for the two classes. Out of the 512 global features, feature number 104 and 410 heave the highest frequency, they are the most important feature for about 24\% and 14\% pairs of samples respectively. Right: The output of the convolution layer for the global feature 104. The plot is generated from about 350 samples for each class whose important feature index is 104.}
    \label{3}
\end{figure}

To do so, we calculate the values of all 512 global features for 20000 samples of the crossover and phase transition events (10000 each). The global feature with largest difference (in the numerical value) is then selected out for each pair of crossover and phase transition events. This feature can be considered an important feature for the given pair of samples. By repeating this for all 10000 pairs, it is possible to find out which global features are the most important global features for most pair of samples. The distribution of the important features (as defined by their feature number from 0 to 511), within the total 10000 pairs of samples, is shown in the left hand side of figure \ref{3}.
It can be seen that for about 2400 pairs of input samples, feature number 104 is an important global feature. The feature map from which feature number 104 is calculated is visualised in the right hand plot of figure \ref{3}. We can see that the values in the feature map distribution are concentrated mostly in two bins, one around -0.8 and another around 1.25. To make clear again, each particle from the input point cloud contributes to some value in the convolutional layer output. In this case most particles either contribute to a value around zero or a value around 1.25. Then, the average is mainly determined by the relative number of particles in the two prominent bins. So, we traced back the particles in both the two bins and investigated their properties separately. We found that all the particles which contributed to the peak on the right in the histogram (at around 1.25) were the fake/empty particles (with zeros for the features) we added into the input data in order to maintain identical input dimensions for all samples. Most of the actual/real particles formed the peak on the left (around -0.8) and very few particles had a value in between the two peaks. Therefore, the global feature-104 is simply a feature that estimates the total track multiplicity in the sample. In other words track multiplicity is one important feature that is used to classify the EoS. However, only because track multiplicity is an important global feature, as learned by the model, does not mean that the multiplicity by itself is sufficient for classification.

\begin{figure}[t]
    \centering
    \includegraphics[width=0.5\textwidth]{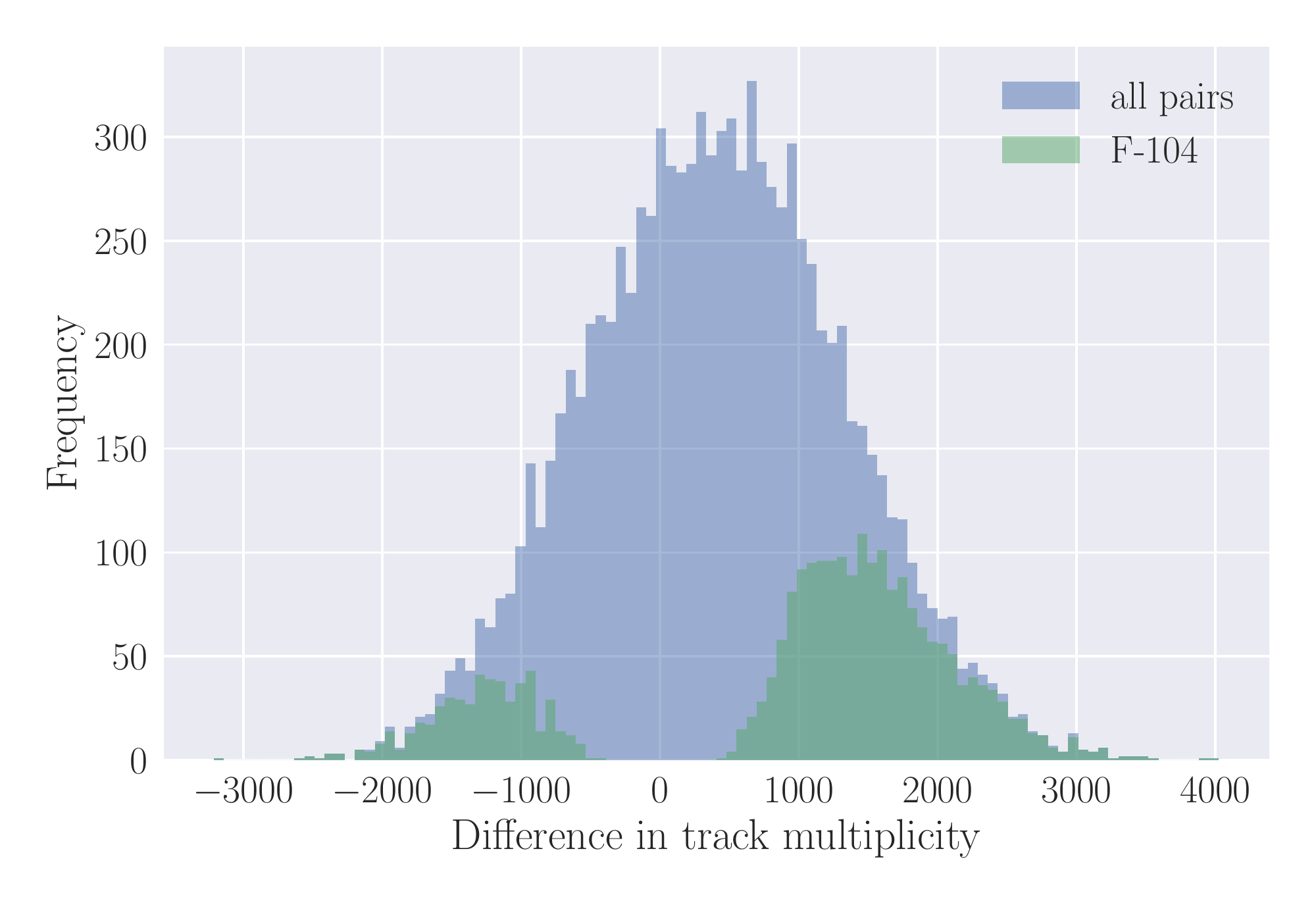}
    \caption{(Color online) Distribution of the track multiplicity difference. The difference in the number of tracks is calculated for 10000 pairs of samples (first-order phase transition - crossover) and plotted in blue colour. The green color depicts the multiplicity difference of all the pairs of samples whose most important Feature is 104 (F-104). Most of the pairs with F-104 contain a significantly large number of tracks for first-order phase transition samples compared to crossover samples.   }
    \label{4}
\end{figure}

In figure \ref{4}, the distribution of the track multiplicity difference (phase transition- crossover) for both the classes are visualised for all pairs of test samples as well as for pairs where feature number 104 (i.e. track multiplicity) is the important feature. It is evident that for the pair of samples whose important feature is feature number 104 (F-104), there is a significant difference in their multiplicities for the two classes. For most of the pairs where that feature is important, phase transition samples contain significantly more particles than a crossover sample with F-104.

On the other hand, for all other events the difference is not well separated which means that using only the difference in multiplicity would not lead to an overall great performance. In other words even though multiplicity can be important, the model does not simply use the track multiplicity to make a decision. The number of tracks is only one of the 512 global features used by the model to classify the data. 
In other words the high accuracy is only obtained by taking into account a multitude of global event features, and their relations, simultaneously. 
Note, that due to the structure of the model it would be very difficult to segregate the samples which can be classified using the track multiplicity and also which other features are taken into account for the final decision. The important feature can also change from pair to pair and we have simply chosen F-104 for an exemplary analysis as it was one of the prominent features. However, it can be seen that other global features also become important features for several other pairs. Therefore it is the combination of all 512 global features that makes it possible to perform an accurate classification. 
Of course one could perform this procedure for all the remaining global features one by one to interpret its physical significance and meaning. For example some of them may be related to the total momentum or momentum differences and correlations. However, this is beyond the scope of this study and is desirable for further investigations in future.

\end{document}